\documentstyle[12pt,psfig]{article}
\renewcommand{\baselinestretch}{1.2}
\newcommand{\beq}{\begin{equation}}
\newcommand{\eeq}{\end{equation}}
\newcommand{\beqa}{\begin{eqnarray}}
\newcommand{\eeqa}{\end{eqnarray}}
\newcommand{\beqar}{\begin{eqnarray*}}
\newcommand{\eeqar}{\end{eqnarray*}}
\newcommand{\eg}{{\it e.g.,}\ }
\newcommand{\ie}{{\it i.e.,}\ }
\newcommand{\labell}[1]{\label{#1}}
\newcommand{\labels}[1]{\label{#1}}
\newcommand{\reef}[1]{(\ref{#1})}
\def\IR{{\hbox{{\rm I}\kern-.2em\hbox{\rm R}}}}
\newcommand{\bth}{{\bar\theta}}
\def\g    {\gamma^2}
\def\st   {\sin ^2\theta}
\def\ct   {\cos ^2\theta}
\def\hf   {\frac{1}{2}}
\def\tT   {\tilde{T}}
\def\L    {\Lambda}
\begin{document}
\def\theequation{\thesection.\arabic{equation}}
\begin{titlepage}
\begin{flushright}
DTP 99-75\\
EHU-FT/9912\\
hep-th/9911007\\
\end{flushright}
\vfil\vfil
\begin{center}

{\Large \bf Composite diholes and intersecting brane--antibrane 
configurations in string/M-theory}

\vspace{25pt}

Auttakit Chattaraputi$^*$\footnote{auttakit.chattaraputi@durham.ac.uk}, 
Roberto Emparan$^{*,\dagger}$\footnote{wtpemgar@lg.ehu.es} 
and 
Anne Taormina$^*$\footnote{anne.taormina@durham.ac.uk}
\vfil

{\sl $^*$Department of Mathematical Sciences}\\
{\sl University of Durham, Durham DH1 3LE, UK}\\
{\sl $^{\dagger}$Departamento de F{\'\i}sica Te\'orica}\\
{\sl Universidad del Pa{\'\i}s Vasco, Apdo. 644, E-48080, Bilbao, Spain}
\vfil

\end{center}

\vspace{5mm}

\begin{abstract}
We construct new configurations of oppositely charged, static black hole
pairs (diholes) in four dimensions which are solutions of low energy
string/M-theory. The black holes are extremal and have four different
charges. We also consider diholes in other theories with an arbitrary
number of abelian gauge fields and scalars, where the black holes can be
regarded as composite objects. We uplift the four-charge solutions to
higher dimensions in order to describe intersecting brane-antibrane
systems in string and M-theory. The properties of the strings and
membranes stretched inbetween these branes and antibranes are studied.
Several other generic features of these solutions are discussed.
\end{abstract}

\vfil\vfil\vfil
\begin{flushleft}
November 1999
\end{flushleft}
\end{titlepage}
\setcounter{footnote}{0}

\newpage
\renewcommand{\baselinestretch}{1.2}  

\section{Introduction}
\renewcommand{\theequation}{1.\arabic{equation}}
\setcounter{equation}{0}

The dual life of D-branes---on the one hand, as objects described by the
conformal field theory of strings that end on them, and on the other
hand as supergravity solitons---is one of the most interesting outcomes
of the so-called ``second string revolution." Indeed, this duality lies
at the heart of the AdS/CFT correspondence \cite{adscft}, and of the
microscopic calculations of black hole entropy \cite{strvaf}. However,
the progress in the construction of exact solutions for self-gravitating
brane configurations lags far behind our understanding of those same
configurations in string perturbation theory. To mention one outstanding
example, we are still far from a satisfactory description of
self-gravitating localized brane intersections, despite much effort and
some progress in certain cases \cite{amanda}. 

Supersymmetric D-branes of type II theories are well-studied examples of
BPS states, which form a very special class of states in the Hilbert
space of string and field theories with extended supersymmetry. So far,
most non-perturbative tests of the duality conjecture are based on such
BPS states, mainly because they are stable, protected from quantum
radiative corrections, and hence, easier to handle. However nice their
properties though, these states do not account for the full spectrum of
any string theory, and the study of non-BPS states, which is still in
its infancy, is essential in any viable attempt at shaping up M-theory.

In the same way as one may interpret BPS D-branes as classical solutions
to type II supergravities, there have been some efforts to construct
supergravity solutions describing a particular class of non BPS states
corresponding to D-brane---anti-D-brane configurations. For instance,
the four-dimensional Kaluza-Klein dipole constructed in
\cite{grossperry} can be embedded in eleven-dimensional supergravity in
order to provide a static $D6$---$\overline{D6}$ brane configuration of
type IIA string theory suspended in an external magnetic field
\cite{dggh2,sen}. More recently, exact solutions to Einstein-Maxwell
theory, with and without dilaton, describing static (but unstable) pairs
of extremal black holes with opposite charges (hereafter, {\it diholes})
were constructed in a background magnetic field, and were
argued to admit an interpretation in terms of a system of intersecting
branes and intersecting antibranes in higher dimensions, after a
suitable uplifting of the four-dimensional solutions when the dilaton
coupling takes one of four special values \cite{blackdih}. The explicit
task of uplifting the solutions in this way has been undertaken since
then in \cite{youm}. Other recent studies of configurations of this type
in the context of string theory include \cite{bert}.

The generalization of these configurations to the case where the charges
of the branes are not equal is the main subject of this paper. After
reviewing in Section 2 the known case of single charge diholes, we
construct in Section 3 a new exact solution of four-dimensional General
Relativity describing oppositely charged, static black hole pairs, where
the black holes are extremal and have an arbitrary number $n$ of
different charges. Black holes of this sort can be regarded as
composites of $n$ extremal, singly charged black holes. Therefore, our
solutions describe composites of $n$ diholes. We however mostly
concentrate on the case $n=4$ with two electric and two magnetic
charges, because of the well-known consistent truncation of a wide class
of low energy superstring theory compactifications to a four-dimensional
action whose bosonic sector,
\begin{eqnarray}
I&=&{1\over 16\pi G}\int d^4x\sqrt{-g}\biggl\{ R-{1\over 2}\left[
(\partial\eta)^2 + (\partial\sigma)^2 +(\partial\rho)^2\right]\nonumber
\\
&-&{e^{-\eta}\over 4}\left[ e^{-\sigma-\rho}F_{(1)}^2
+ e^{-\sigma+\rho}F_{(2)}^2
+e^{\sigma+\rho}F_{(3)}^2 +e^{\sigma-\rho}F_{(4)}^2\right]\biggr\}\ ,
\labell{straction}
\end{eqnarray}
is the starting point of our analysis when considering four abelian
gauge fields. Our construction is inspired by two known results. First
of all, when the branes carry different charges, the corresponding
four-dimensional black holes appear as solutions to theories with four
$U(1)$ gauge fields and three scalars of type \reef{straction}
\cite{cvyo}. Second of all, four-charge pairs of black holes
accelerating apart were found in \cite{rob96,composite}, which, when
adequately written, are very suggestive of the form their static
counterparts might take. (Composites with two charges, which can be
obtained as a particular case of the four-charge case, are also of
interest, see \cite{klop,simon}).

As with other diholes, the composite dihole solutions in an
asymptotically flat space suffer from conical singularities along the
axis of symmetry. These singularities can be removed by suspending the
diholes in external magnetic fields, a procedure we will examine in some
detail. It should be noted that, for reasons to similar those discussed
in \cite{blackdih}, composite diholes are unstable equilibrium
configurations. 

Although these new composite diholes deserve to be studied in their own
right as four-dimensional solutions of General Relativity, they can be
embedded in ten or eleven-dimensional supergravities, and interpreted as
systems of intersecting branes and intersecting anti-branes. This is
discussed in Section 4, in an attempt to test any relevant connections
between supergravity solitons and non BPS states described by
brane---anti-brane type of configurations, and therefore to shed some
light on the duality conjecture beyond the BPS analysis.

\section{Single-charge diholes}\labels{sec:review}
\renewcommand{\theequation}{2.\arabic{equation}}
\setcounter{equation}{0}

We start by reviewing the single-charge dihole solutions as 
they were studied in \cite{blackdih}. Our purpose here is not only to 
introduce 
the structure of the solutions, which is somewhat involved, but also to 
discuss 
the solutions in a coordinatization that is slightly different from the one 
used 
in previous literature, and which will be required later.

It was shown in \cite{bonnor,davged} that the field equations of 
Einstein-Maxwell-Dilaton theory derived from the action
\beq
I={1\over 16\pi G}\int d^4x \; \sqrt{-g}\left(R-2(\partial\phi)^2 
-e^{-2\alpha\phi} F^2\right)\ ,
\labell{emdil}
\eeq
admit solutions with metric
\beqa
ds^2&=&\left( {\Delta+a^2\st\over\Sigma}\right)^{2\over 1+\alpha^2} 
\left[-dt^2 
+ 
{\Sigma^{4\over 1+\alpha^2}\over 
\left(r^2-\g\ct\right)^{3-\alpha^2\over 1+\alpha^2}} 
\left({dr^2\over 
\Delta}+d\theta^2 \right)\right]\nonumber\\
&&+{\Delta 
\sin^2\theta \over \left( {\Delta+a^2\st\over\Sigma}\right)^{2\over 
1+\alpha^2}}d\varphi^2 \ ,
\labell{dilholem}
\eeqa
dilaton 
\beq
e^{-\phi}=\left({\Delta+a^2\st \over\Sigma}\right)^{\alpha\over 1+\alpha^2}\ , 
\labell{dilholed}
\eeq
and (magnetic) one-form gauge potential,
\beq
A={2\over \sqrt{1+\alpha^2}} {M (r+M) a \over 
\Delta +a^2\sin^2\theta}\sin^2\theta\; d\varphi \ ,
\labell{dilholea}
\eeq
 where 
\beqa
\Delta&=&r^2-\g\ ,\nonumber\\
\Sigma&=&(r+M)^2-a^2\cos^2\theta\ .
\labell{defdesi}
\eeqa
The parameter
\beq
\g\equiv M^2+a^2
\eeq
has been introduced here for later convenience. Actually, at fixed value of the dilaton coupling $\alpha$, the two parameters 
defining the solution can be taken to be $\gamma$ and $a$.

When comparing these solutions to those in \cite{blackdih,bonnor,davged} and 
to the literature on similar solutions, it is important to notice a relative 
shift 
of the radial coordinate $r$. To recover the solutions in the form given in 
those papers, let
\beq
r\rightarrow r-M\ .
\labell{rshift}
\eeq

A solution with electric field, dual to the magnetic solution above, can be 
readily constructed. In this case, the dilaton changes from $\phi$ to 
$-\phi$, and the electric gauge potential becomes
\beq
A=-{2\over \sqrt{1+\alpha^2}} {M a\cos\theta \over 
\Sigma} \; dt \ .
\labell{dilholeael}
\eeq

We now follow the analysis in \cite{blackdih}. The solution is clearly 
asymptotically flat as $r\rightarrow\infty$, and in this asymptotic region the 
gauge field is that of a dipole. Although the solution contains apparent 
singularities at $r=\gamma$ (where $\Delta=0$), the actual situation is 
somewhat 
subtle. Notice first that the axis of symmetry of the 
solution (the fixed-point set of the Killing vector $\partial_\varphi$) 
consists 
of the semi-infinite lines $\theta=0,\pi$ (running from $r=\gamma$ to 
$\infty$), and the segment 
$r=\gamma$ that stretches in between them (running from $\theta=0$ to $\pi$). 
The crucial feature of these solutions 
is that at each of the poles $(r=\gamma, \theta=0)$ and $(r=\gamma, 
\theta=\pi)$, lies a
(distorted) extremal charged dilatonic black hole. In order to see this, 
change coordinates $r,\theta$ to $\rho,\bth$,
\beqa
r&=&\gamma +{\rho\over 2} (1+\cos\bth)\ ,\nonumber\\
\sin^2\theta &=&{\rho\over \gamma}(1-\cos\bth)\ ,
\labell{change}
\eeqa
and examine the solution for small values of $\rho$. On doing so, the metric 
in this region takes the form
\beq
ds^2\rightarrow g^{2\over 1+\alpha^2}(\bth)
\left[- \left( {\rho\over q}\right)^{{2\over 
1+\alpha^2}} dt^2 +\left( 
{q\over 
\rho}\right)^{2\over 1+\alpha^2} (d\rho^2 +\rho^2 d\bth^2)\right] +\left( 
{q\over 
\rho}\right)^{2\over 1+\alpha^2}{\rho^2\sin^2\bth \over g^{2\over 
1+\alpha^2}(\bth)}
d\varphi^2 \ ,
\labell{horizon}
\eeq
with
\beq
 g(\bth)=\cos^2(\bth/ 2) +{a^2\over\gamma^2}\sin^2(\bth/ 2)
\ ,
\labell{gbth}
\eeq
and
\beq
q\equiv {M(\gamma +M)\over\gamma}\ .
\labell{defq}
\eeq
For $\alpha=0$ (the case of Einstein-Maxwell theory without a dilaton), this
geometry is that of a Bertotti-Robinson universe ($AdS_2\times S^2$), albeit
distorted by the factor $g(\bth)$.  This is, we find a geometry just like that
of the region close to the horizon ($\rho=0$) of an extremal
Reisnner-Nordstr\"{o}m black hole, but, instead of being spherically
symmetric, it is elongated along the axis in a prolate shape.  For other
values of $\alpha$ the solution at $\rho=0$ has a curvature singularity which
is just like the one at the core of extremal charged dilaton black holes,
although, again, the geometry is not spherically symmetric due to the
distorting factor $g(\bth)$.  Hence we see that the dipolar field of the full
solution is created by two oppositely charged extremal black holes:  a {\it
dihole}\footnote{In \cite{davged} it was argued that for the case $\alpha=1$,
and only for this case, the solution contains regular non-extremal horizons,
and describes a black hole-white hole configuration.  This interpretation is
not consistent with what we have just described:  two extremal horizons,
regular for $\alpha=0$ and singular for $\alpha>0$.}.

That the dipolar field is originated by a pair of extremal black holes, and
not by, say, a pointlike or linear singularity or a pair of charges of a
different kind, is obviously a non trivial issue.  Apparently, Bonnor's dipole
(\ie the $\alpha=0$, non-dilatonic solution found in \cite{bonnor}) was
originally thought to describe a singular pointlike (or segment-like) dipole.
The first identification of a self-gravitating pole-antipole configuration was
made for the case of Kaluza-Klein theory ($\alpha=\sqrt{3}$) in \cite{grossperry}, and
then refined in \cite{dggh2}.  However, in those papers the interpretation was
made on the basis of topological arguments that are particular to the
higher-dimensional structure of Kaluza-Klein theory, and which cannot be
applied to solutions with other values of the dilaton.  The Kaluza-Klein
dipole was later analyzed in \cite{sen} by means of essentially the same
transformation as \reef{change}.  That the solutions \reef{dilholem} of
\cite{davged} actually describe a dihole for arbitrary values of $\alpha$,
including the case $\alpha=0$ which has regular horizons, was first proven in
\cite{blackdih}.

Given the two-black hole interpretation, it would be natural to expect the
dihole solutions to contain the single black hole solutions as the limiting
case where one of the holes is pulled infinitely away from the other.  This is
indeed the case.  Working in coordinates $\rho,\bth$, if the parameter $a$ is
taken to infinity while keeping all other quantities finite, then the solution
reduces precisely to that of a single extremal dilatonic black hole.  The
parameter $a$ plays then the role of a measure of the separation between the
holes. However, this is just a qualitative statement, since the proper spatial
distance between the extremal horizons (for $\alpha=0$) is actually
infinite. A more accurate statement is to say that increasing $a$, while 
keeping the holes' charge fixed, increases the value of the dipole moment (the 
relation, however, becomes approximately linear only for large $a$).

We are primarily interested, however, in the situation where both black holes 
are present in the solution, and therefore we consider finite values of $a$. 
The attraction, gravitational and electromagnetic, that they exert 
on one another is not balanced by any 
external field, so the geometry reacts, as is usual in these situations, by 
producing conical singularities along 
the symmetry axis \cite{blackdih,davged}. 
On physical grounds, it is clear that an external magnetic field aligned with 
the dihole could provide the force to balance the configuration. An exact 
solution containing such a field can be constructed by applying a Harrison 
transformation to \reef{dilholem}. This was done in \cite{blackdih}, and 
results 
in the metric
\beq 
ds^2=\Lambda^{2\over 1+\alpha^2} \left[-dt^2 + {\Sigma^{4\over 
1+\alpha^2}\over 
\left(r^2-\g\ct\right)^{3-\alpha^2\over 1+\alpha^2}} 
\left({dr^2\over 
\Delta}+d\theta^2 \right)\right]+{\Delta 
\sin^2\theta \over \Lambda^{2\over 1+\alpha^2}}d\varphi^2 \ ,
\labell{dilholebm}
\eeq
the dilaton, $e^{-\phi}=\Lambda^{\alpha\over 1+\alpha^2}$, and the gauge 
potential,
\beq
A= {{2\over \sqrt{1+\alpha^2}}M a (r+M)+{1\over 2} 
B[((r+M)^2-a^2)^2+\Delta 
a^2\sin^2\theta] \over 
\Lambda\Sigma}\sin^2\theta\; d\varphi \ ,
\labell{dilholeba}
\eeq
with 
\beqa
\Lambda&=&{1\over \Sigma}\biggl\{ \Delta +a^2\sin^2\theta +2 
\sqrt{1+\alpha^2}B 
M (r+M) a 
\sin^2\theta\nonumber\\
&&+{1+\alpha^2\over 4} B^2 
\sin^2\theta\Bigl[\left((r+M)^2-a^2\right)^2+\Delta a^2\sin^2\theta\Bigr] 
\biggr\} \ ,
\eeqa
and $\Delta$ and $\Sigma$ still given by \reef{defdesi}. At distances much 
larger than the size of the dipole ($r\gg a, M$) the solution asymptotes to 
the 
dilatonic Melvin universe, which describes a self-gravitating, cylindrically 
symmetric magnetic field. It is of interest now to examine what effect the 
external field has on the black hole horizons.
To see this, change coordinates again as in
\reef{change}, and then focus on small values of $\rho$. One finds a geometry 
just like \reef{horizon}, but now the deformation function is
\beq
g(\bth)= \cos^2(\bth/ 2)+\left({a\over\gamma}+B q 
\sqrt{1+\alpha^2}\right)^2\sin^2(\bth/ 2)
\ ,
\labell{fullgbth}
\eeq
$q$ being the parameter defined in \reef{defq}. Notice that if we tune the 
external magnetic field to the value
\beq
B={1\over q\sqrt{1+\alpha^2}}\left( 1-{a\over \gamma}\right)={1\over 
\sqrt{1+\alpha^2}}{\gamma-a\over M(\gamma+M)}\ ,
\labell{equil}
\eeq
then $g(\bth)=1$, and the distortion of the holes disappears\footnote{There is 
a second value of $B$ that yields $g(\bth)=1$, but here we have chosen the one 
for which $B\to 0$ as 
$a\to\infty$.}. It was shown in 
\cite{blackdih} that this very same value of the magnetic field produces a 
cancellation of the conical defects along the symmetry axis, with the choice 
$\triangle\varphi=2\pi$. It is quite 
peculiar that the horizons recover their spherical symmetry precisely when the 
forces in the system are 
balanced and the two black holes are suspended in (unstable) equilibrium in 
the external field. In a 
sense, the latter exactly compensates for the distortions of the horizon 
induced 
by the presence of the other hole.

In the dual electric case the background Melvin field cannot be introduced by 
means of a solution-generating transformation as in the magnetic case. 
Nevertheless, the electric solution can be constructed by straightforward 
dualization of the magnetic one. The dilaton reverses sign as usual, and we 
find 
the electric potential to be
\beqa
A&=&\Biggl[B \cos\theta\left(r-2M +{aBM\sqrt{1+\alpha^2}\over 2} 
(2+\st)\right)\nonumber\\
&&-{2 M a\cos\theta(1-\hf aB\sqrt{1+\alpha^2}\st)^2 \over 
\sqrt{1+\alpha^2}\Sigma}\Biggr]\; dt.
\labell{dilholebat}
\eeqa
This form of the potential manifestly shows how the potential tends to a 
``uniform" field $A_t\rightarrow B r\cos\theta$ as 
$r\rightarrow\infty$. In 
this case $B$ is the asymptotic electric field along the axis.

The physical charge of the holes can be easily read by examining the gauge 
potentials in the region near the horizons. If we keep the field $B$ 
arbitrary, 
instead of fixing it to the equilibrium value \reef{equil}, then, as $\rho\to 
0$, the dilaton goes to 
\beq
e^{-\phi}\to \left({\rho\over q}\; g(\bth)\right)^{\alpha\over 1+\alpha^2}
\eeq
so when the balance condition \reef{equil} is achieved the angular dependence 
disappears. The potential, in its turn, becomes
\beq
A\rightarrow {q\over \sqrt{1+\alpha^2}}\left({a\over \gamma}+ B 
q\sqrt{1+\alpha^2}\right){1-\cos\bth\over g(\bth)}\; d\varphi\ ,
\eeq
or, in the electric case, 
\beq
A\rightarrow -{1\over \sqrt{1+\alpha^2} }\left({a\over \gamma}+ B 
q\sqrt{1+\alpha^2}\right){\rho\over q} 
\; dt\ ,
\eeq
(here we have gauged away a constant) with, of course, a reversal in the sign 
of 
the dilaton. From here we infer 
that the charge is,
\beq
Q={1\over \sqrt{1+\alpha^2}}{q\over{a\over \gamma}+ B 
q\sqrt{1+\alpha^2}}{\triangle\varphi\over 2\pi}\ 
\eeq
in either the electric or magnetic solutions.

\section{Multi-charged diholes}\labels{sec:multi}
\renewcommand{\theequation}{3.\arabic{equation}}
\setcounter{equation}{0}

We address now the construction of new dihole solutions in theories with 
richer 
field content than the single-gauge field theories of \reef{emdil}.

\subsection{String/M-theory diholes with four charges}

In this subsection we consider a theory containing four abelian gauge
fields and three scalars, with action \reef{straction}, which appears as
a consistent truncation of a large variety of compactifications of low
energy string theory, such as toroidally compactified heterotic, IIA and
IIB string theories, and also $D=11$ supergravity
\cite{youm,duff,rahmfeld}. Correspondingly, there is a large number of
possible higher dimensional interpretations of the different gauge
fields and their charge sources. A few of all these possible oxidations
will be discussed in Sec.~\ref{sec:branes}.

Black hole solutions to this theory were constructed in \cite{cvyo}.  The
black holes carry charges $Q_i$, $i=1\dots,4$ under each of the gauge fields,
the charges $Q_1$, $Q_3$ being of magnetic type, and $Q_2$, $Q_4$ electric (or
viceversa, if we consider a dual configuration).  When only $s$ out of the
four possible charges are equal and non-zero, and the rest are zero, then the
theory, and its solutions, reduce to those of the Einstein-Maxwell-dilaton
theory with coupling $\alpha=\sqrt{(4-s)/s}$.  This is, solutions with $1,2,3$
or $4$ equal charges correspond to dilaton coupling
$\alpha=\sqrt{3},1,1/\sqrt{3}$ and $0$, respectively.  The extremal black hole
solutions can be constructed following the ``harmonic function rule" (see \eg
\cite{tseyharm}).  Each gauge field enters in the solution through products of
harmonic functions, in a manner that does  essentially not depend on the other
gauge fields.  In \cite{rob96,composite} it was shown that solutions with two
such black holes accelerating apart could also be found for these
theories (see \cite{simon} for the solutions in a $U(1)^2$ theory).
We are interested here in configurations where the two black holes with
opposite charges are static.

We have managed to construct exact solutions to the field equations for 
these theories with $U(1)^4$ dipole fields. Their metric is
\beqa
ds^2&=&(T_1 T_2 T_3 T_4)^{1/2} 
\left[-dt^2 + \frac{\Sigma_1 \Sigma_2\Sigma_3\Sigma_4}{ 
\left(r^2-\g\ct\right)^3} \left({dr^2\over \Delta}+d\theta^2 
\right)\right]\nonumber\\
&+&\frac{\Delta \sin^2\theta}{(T_1 T_2 T_3 T_4)^{1/2}}
d\varphi^2\ .
\labell{u14m}
\eeqa
Here
\beqa
\Delta&=&r^2-\g\ ,\labell{defdel}\\
\Sigma_i&=&(r+M_i)^2-a_i^2\cos^2\theta\ ,\labell{defsig}\\
T_i&=&\frac{\Delta+a_i^2\st}{\Sigma_i}\ ,\qquad i=1,\dots,4\;.
\labell{defdesii}
\eeqa
The magnetic gauge potentials $A_{(1)}$, $A_{(3)}$,
are given by
\beq
A_{(i)}={2 a_i M_i (r+M_i)\sin^2\theta \over \Delta+a_i^2\sin^2\theta}\;
d\varphi\ ,
\quad i=1,3
\labell{u14amag}
\eeq
whereas the electric potentials $A_{(2)}$, $A_{(4)}$, are
\beq
A_{(i)}=-{2 a_i M_i \cos\theta \over \Sigma_i}\;dt\ ,
\quad i=2,4\ .
\labell{u14aele}
\eeq
The scalar fields, in turn, take the form
\beq
e^{-2\eta} = \frac{T_1 T_3}{T_2 T_4},  \qquad
e^{-2\sigma} = \frac{T_1 T_4}{T_2 T_3},  \qquad
e^{-2\rho} = \frac{T_1 T_2}{T_3 T_4}.
\labell{u14sc}
\eeq
The solutions are parameterized in terms of five independent parameters. 
Physically, the parameters can be regarded as fixing the four charges of the 
holes and the 
``separation" between the pair. In practice, we will choose the independent 
parameters to be $\gamma$ (which we take to be positive) and all the $a_i$ 
(satisfying $|a_i|\leq \gamma$). The other parameters $M_i$ are not 
independent, 
but rather given by
\beq
M_i^2=\gamma^2-a_i^2\ .
\eeq
When $a_i^2=\g$ for all $i$, then all the $M_i$'s vanish and the metric is 
that of flat space. In general we can have some $M_i=0$ for some values of $i$, 
and nonzero for others, and get a non-trivial solution.

There are several non-obvious aspects in going from the solutions of the
single-gauge field theory to the solutions of the $U(1)^4$ theory.  One of them
is that the combination $a_i^2+M_i^2$ should take the same value for all $i$,
so with our choice of radial coordinate, the function $\Delta$ is the same for all values of $i$.
Another point is related to the characteristic way in which the metric functions
in \reef{u14m} factorize into contributions from each separate gauge field 
(a similar factorization had been observed also for Melvin fields and
accelerating black holes in \cite{composite}). The way the factorization
happens in these solutions relies crucially on our choice of the radial
coordinate, explained in the previous section. To see this, realize that
when more than one parameter $M_i$ is involved, the radial shift
\reef{rshift} cannot be properly undone\footnote{Notice that this
implies that the ``modified harmonic function" rule conjectured in
\cite{youm} on the basis of the solutions with a single gauge field is
of little use in trying to get to the new solutions with different
charges, since that rule was based on a choice of radial coordinate that
is inappropriate for this purpose.}. The factorization suggests that the
$U(1)^4$ dihole can then be thought of as a composite of four diholes.
For an isolated $U(1)^4$ black hole it is possible to separate, at zero
cost in energy, each of the constituents from the other three, \ie the
black hole can be regarded as a composite of four marginally bound
components \cite{rahmfeld}. However, it is not clear whether we can
separate, at zero energy cost, the single-charge component diholes of a
composite dihole. It may well be that what was in isolation a state
bound at threshold (the four- charge extremal black hole) becomes
non-marginally bound in the presence of its anti-state. 

It is also a straightforward matter to check that the dilatonic dihole solutions
\reef{dilholem} for $\alpha=\sqrt{3},1,1/\sqrt{3}$ and $0$ are recovered by
taking $1,2,3$ or $4$ non-zero and equal values of $M_i$.  

The analysis of these solutions can be done in exactly the same manner as we
have done for Bonnor's dipole and its dilatonic counterparts.  Coordinate
singularities occur when $r=\gamma$, and these turn out to be, away from the
poles, conical singularities.  Again, a straightforward analysis of the
conical deficits along the various portions of the symmetry axis reveals that
it is not possible to eliminate the deficit along the segment $r=\gamma$ with
the natural choice of period $\triangle\varphi=2\pi$ which cancels the deficit
along the lines $\theta=0,\pi$.  However, these singularities can be resolved
by introducing magnetic background fields in our axisymmetric solutions by
means of the generalized Harrison transformation constructed in \cite{composite}
for the $U(1)^4$ theory, and by subsequently tuning them to a value which
eliminates the conical deficit.  The latter point, we will see, becomes
somewhat subtle when more than one gauge field is present.

After applying the generalized Harrison transformation, the metric of the 
$U(1)^4$ dipole
solution becomes
\beqa
ds^2&=&(\Lambda_1 \Lambda_2 \Lambda_3 \Lambda_4)^{1/2} 
\left[-dt^2 + \frac{\Sigma_1 \Sigma_2\Sigma_3\Sigma_4}{ 
\left(r^2-\g\ct\right)^3} \left({dr^2\over \Delta}+d\theta^2 
\right)\right]\nonumber\\
&+&\frac{\Delta \sin^2\theta}{(\Lambda_1 \Lambda_2 \Lambda_3 \Lambda_4)^{1/2}}
d\varphi^2\ ,
\labell{hu14m}
\eeqa
the magnetic and electric potentials are given by,
\beq
A_{(i)}= { 2 a_i M_i (r+M_i)\sin^2\theta + \hf B_i\st 
                \left[ \left( \left( r+M_i \right)^2-a_i^2 \right)^2+
                       a_i^2 \Delta\st \right] \over
                \Lambda_i\Sigma_i}\;d\varphi
\;,\quad i=1,3
\labell{hu14amg}\eeq
\beqa
A_{(i)}&=& \Biggl[ B_i \cos\theta\left(r-2M_i +{a_iB_iM_i\over 2} 
(2+\st)\right) \nonumber\\
&&-{2 M_i a_i\cos\theta(1-\hf a_iB_i\st)^2 \over 
\Sigma_i}\Biggr]\;dt
\;,\quad i=2,4
\labell{hu14ael}
\eeqa
and the scalars by,
\beq
e^{-2\eta}={\Lambda_1\Lambda_3\over\Lambda_2\Lambda_4} \ ,\qquad 
e^{-2\sigma}={\Lambda_1\Lambda_4\over\Lambda_2\Lambda_3} \ ,\qquad 
e^{-2\rho}={\Lambda_1\Lambda_2\over\Lambda_3\Lambda_4} \ ,
\eeq
where
\beq
\Lambda_i=
{\Delta + a_i^2\st+2B_ia_iM_i(r+M_i)\st+\frac{1}{4}B_i^2 \st 
  \left[ \left( \left( r+M_i \right)^2-a_i^2 \right)^2+
                       a_i^2 \Delta\st \right] \over
  \Sigma_i}\ .
\labell{hu14la}
\eeq
This solution obviously reduces to the previous one if we set $B_i=0$. We have 
denoted the external fields collectively as $B_i$, even if for $i=2,4$ 
they are electric fields. Observe that the metric and scalars can be obtained 
from \reef{u14m} and \reef{u14sc} by simply substituting $\Lambda_i$ for 
$T_i$. 

Along the outer semi-axes $\theta=0,\pi$ the conical deficit is given by 
$\delta_{(\theta=0,\pi)}=2\pi-\triangle\varphi$, no matter what the value of 
the external fields $B_i$ is. We thus choose
$\triangle\varphi= 2\pi$ in order to remove the conical deficit on that
portion of the symmetry axis. On the other hand, the deficit along the inner
segment of the axis, $r=\gamma$, is calculated to be,
\beq
\delta_{(r=\gamma)}=
2\pi-\prod_{i=1}^4 \left( {a_i \over \gamma}+B_i q_i \right)^{-1}
\triangle\varphi , 
\eeq
where we have defined 
\beq
q_i\equiv {M_i(\gamma+M_i) \over \gamma}\ .
\labell{charges}\eeq
We can see that, for the choice of period in the variable $\varphi$ we 
made 
earlier, \ie 
$\triangle \varphi= 2\pi$, the conical 
singularity along $r=\gamma$ disappears when 
\beq
\prod_{i=1}^4 \left( {a_i \over \gamma}+B_i q_i \right)=1.
\labell{u14balance}
\eeq
A {\it very particular}
solution of this equation is obtained by requiring each of the factors on the 
l.h.s.\ to be equal to $1$, 
\ie 
set the strength of the external fields to the values
\beq
B_i={\gamma-a_i \over M_i(\gamma+M_i)}
   ={2M_i \over (\gamma+M_i + a_i)^2}\ ,\qquad i=1\dots,4
\labell{bivalue}
\eeq
which can be interpreted as a separate force-balance condition for each of the 
gauge fields. Nevertheless, it should be kept in mind that \reef{bivalue} is 
by 
no means a typical solution. On the contrary, the balance of forces will 
typically be achieved with different contributions from each factor in 
\reef{u14balance}. As a matter of fact, it is even possible to satisfy 
\reef{u14balance} for diholes with four different charges by turning on only 
one 
external field, say $B_1$.

Although the metric still appears to be singular at the endpoints of the 
$U(1)^4$ dipole, one can actually reveal its true structure by studying the 
solution near the throat as we did in the previous section. As before, we may 
explore this 
region
by changing coordinates from $(r,\theta)$ to $(\rho, \bth)$ as in 
\reef{change} 
and by
keeping $\rho$ much smaller than any other scale in the problem.
Near $r=\gamma, \theta =0$, the metric becomes\footnote{We are assuming here 
that all four charges are non-zero. The modifications for the case where some 
of 
them vanish can be inferred easily.},
\beq
ds^2 =g^2(\bth)\left[ -{\rho^2\over q^2} dt^2 + {q^2\over \rho^2} d\rho^2 +q^2 
d\bth^2 \right]
+{q^2\sin^2\bth\over g^2(\bth)} d\varphi^2 \ ,
\labell{deformthroat4}
\eeq
where
$q=(q_1 q_2 q_3 q_4)^{1/4}$,
and where $g(\bth)=\left[g_1(\bth) g_2(\bth) g_3(\bth) g_4(\bth)\right]^{1/4}$ 
with
\beq
g_i(\bth)= \cos^2(\bth/ 2)+\left({a_i\over\gamma}+B_iq_i\right)^2
\sin^2(\bth/ 2)
\labell{deform4}
\eeq
a function such that $g_i(\bth)=1$ when the field $B_i$ is tuned to the value 
\reef{bivalue}. However, for more generic cases the deformation function 
$g(\bth)$ will be different from $1$ even if the conical singularities are 
cancelled through \reef{u14balance}, and the horizons will in general be 
deformed. This feature is particular to theories with more than one gauge 
field. In any event, we see that near the poles the solution exhibits, apart 
from the distortion, the same structure as a four-charge black hole near its 
horizon.

The gauge fields are also distorted near the throat, where they are given by
\beqa
A_{(i)}&=&{q_i \over g_i(\bth)}\left( {a_i\over\gamma}+B_iq_i\right) 
(1-\cos\bth )\;d\varphi \;, \quad i=1,3\nonumber\\
A_{(i)}&=&-{\rho\over q_i}\left({a_i\over \gamma}+ B_i q_i\right)\;dt \;, 
\quad 
i=2,4
\labell{deformpole4}
\eeqa
and the corresponding physical charges are
\beq
Q_i={\triangle\varphi\over 2\pi}{q_i\over {a_i\over\gamma}+B_iq_i} \ .
\labell{phcharge}\eeq
Finally, the scalar fields in this limiting region are
\beq
e^{-2\eta}={q_2 q_4\over q_1 q_3}{g_1(\bth)g_3(\bth)\over g_2(\bth)g_4(\bth)} 
\;,\quad 
e^{-2\sigma}={q_2 q_3\over q_1 q_4}{g_1(\bth)g_4(\bth)\over 
g_2(\bth)g_3(\bth)}
\;,\quad 
e^{-2\rho}= {q_3 q_4\over q_1 q_2}{g_1(\bth)g_2(\bth)\over 
g_3(\bth)g_4(\bth)}\;,
\eeq
which present the unusual feature that, in general, they will vary as we move 
along the horizon.

The deformation of the black hole horizons allows us to check a non-trivial 
aspect of the 
entropy-area law for black holes. Notice that when all four charges are 
turned on, the black holes have a non-singular, deformed horizon with 
non-vanishing area. Now, for an {\it isolated} extremal black hole the area is 
entirely determined by its physical charges $ Q_i$ as ${\cal 
A}_h=4\pi\sqrt{Q_1Q_2Q_3Q_4}$. 
This area can be associated, through the Bekenstein-Hawking 
law, with an entropy. On physical grounds we would expect the entropy of 
the system to remain unchanged if its physical charges, which fix the state, 
remain fixed, no matter what the distortion of the horizon may be. It is by no 
means clear that the solutions given above should satisfy this property. 
Nonetheless, the area of each of the horizons in the dihole configuration is
\beq
{\cal A}_h=4\pi \sqrt{q_1q_2q_3q_4}=4\pi \sqrt{Q_1Q_2Q_3Q_4}\ ,
\eeq
where the last equality is obtained when the singularities are cancelled by 
requiring \reef{u14balance} (but not necessarily \reef{bivalue}) and 
$\triangle\varphi=2\pi$. Hence, the area as a 
function of the physical charges remains unaltered, despite the deformation of 
the black hole horizon. A similar test of the invariance of the entropy under 
deformations of the horizon was performed in \cite{composite}.

Finally, it is a straightforward exercise to show that when one of the holes
is pulled away by making $\gamma$ large, while keeping $r-\gamma$, $\gamma\st$
and $M_i$ finite, we get $\Sigma_i \simeq 2\gamma(\rho+M_i)$, $q_i \simeq
M_i\simeq Q_i$ and $\Lambda_i \simeq (1+{Q_i \over \rho})^{-1}$, so that the
metric becomes that of an isolated extremal $U(1)^4$ black hole.

\subsection{$U(1)^n$ composite diholes}

The results we have just described can be generalized to the following 
theories 
containing $n$ gauge 
fields and $n-1$ independent scalars with action
\begin{equation}
I={1\over 16\pi G}\int d^4x\sqrt{-g}\biggl\{ R-{1\over 2 n^2}
\sum_{i=1}^n\sum_{j=i+1}^n(\partial\sigma_i-\partial\sigma_j)^2
-{1\over n}\sum_{i=1}^n
e^{-\sigma_i} F_{(i)}^2\biggr\},
\labell{unaction}
\end{equation}
and with the scalars satisfying
\begin{equation}
\sum_{i=1}^n\sigma_i =0.
\labell{singular}
\end{equation}
Such theories were considered in \cite{composite} as a generalization of the 
theories with four abelian gauge fields we have just considered. In general, 
these $U(1)^n$ 
theories do not seem to be related 
to low energy string/M-theory, nor to any other supergravity theory in four 
dimensions. Nevertheless they exhibit the same peculiarities as the $U(1)^4$ 
theories, which  are merely a particular case of the above type of theory, 
as described in \cite{composite}. All these theories admit black 
hole solutions which follow the ``harmonic function rule", as well as 
solutions 
with two black holes accelerating apart \cite{composite}. It is
therefore natural to 
expect that dihole solutions can be constructed as well. Indeed, their metric 
is
\beq
ds^2=\prod_{i=1}^n T_i^{2/n} 
\left[-dt^2 + \frac{\prod_{i=1}^n \Sigma_i^{4/n}}{ 
\left(r^2-\g\ct\right)^3} \left({dr^2\over \Delta}+d\theta^2 
\right)\right]
+\frac{\Delta \sin^2\theta}{\prod_{i=1}^nT_i^{2/n}}
d\varphi^2\ ,
\labell{u1nm}
\eeq
where $\Delta$, $\Sigma_i$ and $T_i$ take the same form as in \reef{defdel},\reef{defsig},\reef{defdesii}.
The potentials (in magnetic form) are
\beq
A_{(i)}={2 a_i M_i (r+M_i)\sin^2\theta \over \Delta+a_i^2\sin^2\theta}\;
d\varphi\ ,
\qquad i=1,\dots,n
\labell{u1na}
\eeq
and the scalars
\beq
e^{-\sigma_i}= T_i^2
\prod_{j=1}^n T_j^{-2/n}.
\labell{u1nsc}
\eeq

The qualitative features of these solutions are precisely the same as for the 
$U(1)^4$ case of the previous subsection, so our discussion  will be 
rather cursory.
These solutions have $n+1$ independent parameters, $\{\gamma, 
a_i,i=1,\dots,n\}$, 
while $M_i$ are fixed by $M_i^2=\gamma^2-a_i^2$.
By setting $n=1$ and shifting the coordinate $r$ to $r-M_1$,
one recovers Bonnor's magnetic dipole solution of non-dilatonic
Einstein-Maxwell theory \cite{bonnor}. The dilatonic solutions of 
\cite{davged} 
can also be recovered in a simple manner for rational values of $\alpha^2$ 
\cite{composite}. To this effect, take $s$ out of the $n$ possible
parameters 
$M_i$ 
(say, $i=1,\dots,s$) to be equal and non-zero, and the remaining $M_i$, 
$i=s+1,\dots,n$ to be vanishing. In this way, the solutions \reef{dilholem} 
are 
recovered, with 
dilaton coupling $\alpha=\sqrt{n/s-1}$, and with the fields identified as
\beqa
\sigma_1&=&\dots=\sigma_s=2\alpha\phi\ ,\qquad 
\sigma_{s+1}=\dots=\sigma_n=-{2\over \alpha}\phi\ ,\nonumber\\
F_{(i)}&=&\sqrt{\alpha^2+1}\; F\ ,\quad i=1,\dots,s\ .
\eeqa

The conical singularities that the solutions possess can be removed by means
of the generalized Harrison transformation for the $U(1)^n$ theory
constructed in \cite{composite}, and by
subsequently tuning them to a value which eliminates the conical deficit.

After applying the generalized Harrison transformation, the $U(1)^n$ dipole
solution becomes,
\beq
ds^2=\prod_{i=1}^n \Lambda_i^{2/n} \left[-dt^2 + \frac{\prod_{i=1}^n 
\Sigma_i^{4/n}}{ 
\left(r^2-\g\ct\right)^3} \left({dr^2\over \Delta}+d\theta^2 
\right)\right]+\frac{\Delta \sin^2\theta}{\prod_{i=1}^n \Lambda_i^{2/n}}
d\varphi^2 ,
\labell{hu1nm}
\eeq
\beq
e^{-\sigma_i}=\Lambda_i^2 \prod_{j=1}^n \Lambda_j^{-2/n},
\eeq
where the magnetic gauge potentials and $\Lambda_i$ are given in \reef{hu14amg} 
and \reef{hu14la}.

All the features of the $U(1)^4$ solutions we described in the previous 
subsection carry over to the generic $U(1)^n$ case modulo some obvious 
adjustments.

\section{Intersecting brane-antibrane configurations}\labels{sec:branes}
\renewcommand{\theequation}{4.\arabic{equation}}
\setcounter{equation}{0}

We already pointed out that the solutions to the $U(1)^4$ theory admit 
embeddings into higher dimensional supergravity theories arising from 
string/M-theory at low energies. When uplifted to $D \ge 5$ dimensions, each  individual charge \reef{phcharge} will typically be interpreted as the charge (density) of a $p$-brane, with all its 
spatial directions wrapped around a $p$-torus, and delocalized in $D-p-4$ 
transverse directions. We are referring to branes in a manner loose enough to 
allow for pp-waves and KK monopoles to be introduced in a straightforward way 
in the discussion  
\footnote{More precisely, any charge that would naturally enter a solution 
through a harmonic function can be paired up with its anticharge to provide
a characterization of brane-antibrane systems}.

\begin{figure}[ht]
\hskip1cm
\psfig{figure=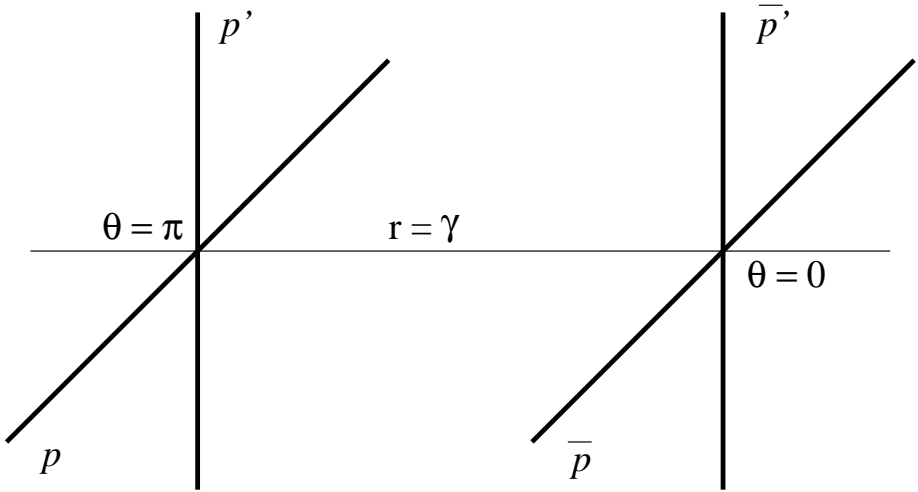}
\hskip2cm
\caption{Geometry of the $p-p'$ brane and $\bar p-\bar p'$ brane
intersections. The direction labelled $p$ denotes directions along the
$p$-brane which are transverse to the $p'$-brane, and viceversa. The
same applies, mutatis mutandis, for the `anti-intersection' $\bar p-\bar
p'$. The $p$- and $\bar p$-branes are parallel. In the solutions
described in the text the branes are delocalized in their relative
transverse directions, and also in all but three overall transverse
dimensions ($r,\theta,\varphi$). Of the latter directions, only the
symmetry axis is shown in the figure. When infinitely separated from the
other intersection ($\gamma\rightarrow\infty$), the $p$- and $p'$-branes
we consider are marginally bound to each other.} 
\label{fig:ppbar} 
\end{figure}

Solutions with just one out of four non-vanishing charges correspond to
single brane solutions, whereas solutions with more than one charge
describe brane intersections or (marginal) bound states of branes. A
$U(1)^4$ dihole solution, when viewed in this way, can be oxidized to
describe an intersection of up to four branes and an `anti-intersection'
of the corresponding anti-branes. We have tried to sketch such a
configuration in Fig.~\ref{fig:ppbar}. Each brane is parallel to its
anti-brane, and the whole system is delocalized in such a way that the
branes are localized in the overall transverse directions only.

The construction of brane-antibrane solutions based on the $U(1)^4$
dihole solutions described in Sec.~\ref{sec:multi} does not
significantly differ in its concept from the way a one-black hole
solution with four or less charges is uplifted to a brane configuration.
Indeed, it was pointed out in \cite{blackdih} (see also \cite{youm})
that the diholes of Einstein-Maxwell-dilaton theory could be uplifted in
a straightforward way to intersecting brane-anti-brane configurations of
the sort just described, with the severe restriction that the charges of
the intersecting branes should all be equal. This restriction can be
relaxed when using our new $U(1)^4$ solutions, which allow for a richer
catalog of configurations. The factorized form of the solutions,
particularly that of the scalar fields involved, greatly helps in
deducing the form of the higher dimensional (internal) metric components
from those of ordinary brane configurations: the harmonic functions
$H_i$ of the latter get replaced by the functions $T_i^{-1}$ or
$\Lambda_i^{-1}$, whose inverses were introduced in \reef{defdesii} and
\reef{hu14la}. We stress that this rule, however, applies only to the
internal dimensions and not to the four-dimensional part of the
solution.

Note that we are unable to consider non-extremal branes, since a
solution describing a pair of non-extremal charged black holes is not
available. Let us also emphasize that solutions {\it without} an
external field contain conical singularities along the symmetry axis,
and a physical interpretation in string/M-theory in terms of, \eg local
cosmic strings, is not clear. Nevertheless, it is possible to
remove these singularities by introducing an external field, using a
similar procedure to the one described in the previous sections. 

\subsection{Some explicit examples}

We now illustrate the uplifting of our $U(1)^4$ dihole solutions to
three different brane-antibrane configurations.
\\

(1) The $3\perp 3\perp 3\perp 3$ and $\bar 3\perp\bar 3\perp\bar
3\perp\bar 3$ system in $D=10$ type IIB theory.

Let us consider the ten-dimensional metric,
\begin{eqnarray}
ds_{10}^2= \left( T_1 T_2 T_3 T_4\right)^{1/2} \left[-dt^2 + \frac{ \Sigma_1
\Sigma_2 \Sigma_3 \Sigma_4}{
\left(r^2-\g\ct\right)^3} \left({dr^2\over \Delta}+d\theta^2 
\right)\right]+\frac{\Delta \sin^2\theta}{\left(T_1 T_2 T_3 T_4\right)^{1/2}}
d\varphi^2 \nonumber\\
+ \left(\frac{T_1T_2}{T_3T_4}\right)^{1/2}dx_1^2 + 
\left(\frac{T_1T_3}{T_2T_4}\right)^{1/2}dx_2^2
+ \left(\frac{T_1T_4}{T_2T_3}\right)^{1/2}dx_3^2 \nonumber\\
+ \left(\frac{T_2T_3}{T_1T_4}\right)^{1/2}dx_4^2+ 
\left(\frac{T_2T_4}{T_1T_3}\right)^{1/2}dx_5^2
+ \left(\frac{T_3T_4}{T_1T_2}\right)^{1/2}dx_6^2.
\labell{uplift1}\end{eqnarray}
with the functions $\Sigma_i$ and $T_i$ defined in \reef{defsig}, \reef{defdesii}. The
ten-dimensional dilaton is constant,
and the five-form field strength is given by,
\begin{eqnarray}
F_{[5]}= dA_1\land dx_4\land dx_5\land dx_6+dA_2\land dx_1\land dx_4\land dx_5 
\nonumber\\
 +dA_3\land dx_1\land dx_3\land dx_5+dA_4\land dx_3\land dx_5\land dx_6.  
 \labell{5form1}\end{eqnarray}
with the magnetic potentials $A_{1/3}$ given in \reef{u14amag} and the electric 
potentials $A_{2/4}$ given in \reef{u14aele}. We also use $r^2 = x_7^2+x_8^2+x_9^2$.
The compactification of the above type IIB solution on a six-dimensional torus
yields the $U(1)^4$ dipole solution \reef{u14m}.
To check that the system indeed contains brane-antibrane pairs, one may
change coordinates from $(r,\theta)$ to 
$(\rho,\bar\theta)$ as in \reef{change}. In the limit where the parameter 
$a_i$ is large, and where $\theta \rightarrow 0$, the function $T_i$ becomes 
the inverse of the harmonic function 
of a delocalized D3-brane
$ T_i \to \tilde{T}_i=\left(1+\frac{Q_i}{\rho}\right)^{-1}$, and the metric becomes
\begin{eqnarray}
ds_{10}^2= 
\left( \tT_1 \tT_2 \tT_3 \tT_4\right)^{1/2} (-dt^2) 
+ \left(\tT_1\tT_2\tT_3\tT_4\right)^{-1/2}
\left(d\rho^2+\rho^2\left(d\bar\theta^2 
+\sin^2\bar\theta d\varphi^2\right) \right) \nonumber\\
+ \left(\frac{\tT_1\tT_2}{\tT_3\tT_4}\right)^{1/2}dx_1^2 + 
\left(\frac{\tT_1\tT_3}{\tT_2\tT_4}\right)^{1/2}dx_2^2
+ \left(\frac{\tT_1\tT_4}{\tT_2\tT_3}\right)^{1/2}dx_3^2 \nonumber\\
+ \left(\frac{\tT_2\tT_3}{\tT_1\tT_4}\right)^{1/2}dx_4^2+ 
\left(\frac{\tT_2\tT_4}{\tT_1\tT_3}\right)^{1/2}dx_5^2
+ \left(\frac{\tT_3\tT_4}{\tT_1\tT_2}\right)^{1/2}dx_6^2.
\labell{4D3}\end{eqnarray}
This is exactly the solution for the four D3-brane intersection constructed in
\cite{laba,tseykle}. The solution for four anti-D3-branes intersection
(i.e. for four D3-branes with opposite charge) is obtained when
taking the $\theta \to \pi$ limit instead of the $\theta \rightarrow 0$
limit. However, in order to show that the system consists of the
intersection of four D3-branes together with the intersection of four
$\overline{D3}$-branes in ten dimensions, one must consider finite
values of $a_i$ and take the near horizon limit of \reef{uplift1} in
complete analogy with the four-dimensional case. As already mentioned
earlier, this solution has a conical deficit along the symmetry axis, 
which pulls the branes apart from each other. However, as can be anticipated
from the discussion of $U(1)^n$ diholes suspended in external magnetic fields,
such a brane/anti-brane configuration can be cured of any conical deficit
along the symmetry axis by tuning the magnetic field $B$ to an appropriate value. The relevant ten-dimensional metric is then given by \reef{uplift1},
where the four functions $T_i$ are replaced by the functions $\Lambda_i$
given in \reef{hu14la}. The five-form field-strength is again formally written as in \reef{5form1}, with the magnetic potentials $A_{1/3}$ and electric potentials $A_{2/4}$ given by \reef{hu14amg} and \reef{hu14ael} respectively.
In the near horizon limit, and with $a_i$ large $(r\gg M_i)$, the
functions $\Lambda_i$, for $\theta \rightarrow 0$, become the harmonic
functions
$\tT_i$ involved in the description of the four D3-brane intersections, and
the limiting metric is \reef{4D3}. If $\theta \rightarrow \pi$ instead,
one obtains the four $\overline{D3}$-brane intersections. One concludes
that the metric  
\begin{eqnarray}
ds_{10}^2= \left( \L_1 \L_2 \L_3 \L_4\right)^{1/2} 
\left[-dt^2 + \frac{ \Sigma_1 \Sigma_2 \Sigma_3 \Sigma_4}{ 
\left(r^2-\g\ct\right)^3} \left({dr^2\over \Delta}+d\theta^2 
\right)\right]+\frac{\Delta \sin^2\theta}{\left(\L_1 \L_2\L_3\L_4\right)^{1/2}}
d\varphi^2 \nonumber\\
+ \left(\frac{\L_1\L_2}{\L_3\L_4}\right)^{1/2}dx_1^2 + 
\left(\frac{\L_1\L_3}{\L_2\L_4}\right)^{1/2}dx_2^2
+ \left(\frac{\L_1\L_4}{\L_2\L_3}\right)^{1/2}dx_3^2 \nonumber\\
+ \left(\frac{\L_2\L_3}{\L_1\L_4}\right)^{1/2}dx_4^2+ 
\left(\frac{\L_2\L_4}{\L_1\L_3}\right)^{1/2}dx_5^2
+ \left(\frac{\L_3\L_4}{\L_1\L_2}\right)^{1/2}dx_6^2.
\end{eqnarray}
is that of a system made of the intersection of four D3-branes and of the
intersection of four $\overline{D3}$-branes.
\\

(2) The $2\perp 2\perp 5\perp 5$ and $\bar 2\perp\bar 2\perp\bar 5\perp\bar 5$ 
system in $D=11$ supergravity.

This system may be described by the $D=11$ metric,

\begin{eqnarray}
ds_{11}^2= \left( T_2 T_4\right)^{2/3}\left( T_1 T_3\right)^{1/3} \left[-dt^2 
+ 
\frac{ \Sigma_1\Sigma_2\Sigma_3 \Sigma_4}{\left(r^2-\g\ct\right)^3} 
\left({dr^2\over \Delta}+d\theta^2 
\right)\right] \nonumber\\
+\frac{\Delta \sin^2\theta}{\left( T_2 T_4\right)^{1/3}\left(T_1 
T_3\right)^{2/3}}d\varphi^2 
+ \left(\frac{T_2^2T_1}{T_4T_3^2}\right)^{1/3}dx_1^2 + 
\left(\frac{T_2^2T_3}{T_4T_1^2}\right)^{1/3}dx_2^2 \nonumber\\
+ \left(\frac{T_4^2T_1}{T_2T_3^2}\right)^{1/3}dx_3^2 
+ \left(\frac{T_4^2T_3}{T_2T_1^2}\right)^{1/3}dx_4^2
+
\left(\frac{T_1T_3}{T_2T_4}\right)^{1/3}\left(dx_5^2+dx_6^2+dx_7^2\right),
\labell{uplift2}\end{eqnarray}
with four-form field strength,
\begin{eqnarray}
F_{[4]}= 3dA_1\land dx_2\land dx_4-3dA_2\land dx_1\land dx_2\nonumber\\
 +3dA_3\land dx_1\land dx_3-3dA_4\land dx_3\land dx_4.
\end{eqnarray}
The magnetic potentials $A_{1/3}$ are again defined by \reef{u14amag}  and the electric potentials 
$A_{2/4}$ by \reef{u14aele}. Here,  $r^2 = x_8^2+x_9^2+x_{10}^2$. 
In the $(\rho,\bar\theta)$ coordinate system, and in the  
limit of $a_i$ large 
and $\theta \to 0$,  $T_{2/4}^{-1}$ and $T_{1/3}^{-1}$ become the 
harmonic functions required for the description of electric M2 branes and magnetic M5 branes.
The solution describes the system of intersecting branes (two M2 and two M5)
and intersecting anti-branes ( two $\overline {\rm M2}$ branes 
and two $\overline {\rm M5}$-branes), and becomes the dipole solution in 
$U(1)^4$ theory when compactified on a seven-torus.
Once again, we may construct an analog solution in the presence of an external
magnetic field in order to remove any conical singularity arising in the
above solution. Replacing $T_i$ by $\Lambda_i$ everywhere in \reef{uplift2}, and
taking the potentials to be \reef{hu14amg} and \reef{hu14ael}, one obtains 
a solution to $D=11$ supergravity which again describes a system of
intersecting branes (two M2 and two M5)
and intersecting anti-branes ( two $\overline {\rm M2}$ branes 
and two $\overline {\rm M5}$-branes), but this time with non-zero magnetic
field $B$. \\

(3) The $2\perp 2\perp 2\perp 6$ and $\bar 2\perp\bar 2\perp\bar
2\perp\bar 6$ system in $D=10$ type IIA theory.

We end this subsection with a configuration we can relate to the
$D6$---$\overline{D6}$ system studied by Sen, and which we will employ
later in order to characterize the string stretching
between the branes and anti-branes using arguments similar to those in
\cite{sen}.

The configuration is described by a solution to type IIA supergravity
with metric
\begin{eqnarray}
ds_{10}^2=  T_1^{1/8} \left(T_2 T_3 T_4\right)^{5/8} \left[-dt^2 + \frac{ \Sigma_1\Sigma_2 \Sigma_3 \Sigma_4}{\left(r^2-\g\ct\right)^3} 
\left({dr^2\over \Delta}+d\theta^2 \right)\right]
+\frac{\Delta \sin^2\theta}{T_1^{7/8}\left( T_2 T_3 T_4\right)^{3/8}}
d\varphi^2 \nonumber\\
+ \frac{T_1^{1/8}T_2^{5/8}}{T_3^{3/8}T_4^{3/8}}(dx_1^2 + dx_2^2)
+ \frac{T_1^{1/8}T_3^{5/8}}{T_2^{3/8}T_4^{3/8}}(dx_3^2 +dx_4^2) 
+ \frac{T_1^{1/8}T_4^{5/8}}{T_2^{3/8}T_3^{3/8}}(dx_5^2+dx_6^2)
\labell{uplift3}\end{eqnarray}
and dilaton
\beq
e^{4\phi}={T_1^3\over T_2 T_3 T_4},
\eeq
with the functions $\Sigma_i$ and $T_i$ defined in \reef{defsig},\reef{defdesii}.
Here, $T_1$ is associated with the $D6$ brane while $T_i, i=2,3,4$ are
associated with the three $D2$ branes. We first discuss in which context
this metric corresponds to a $D6$---$\overline{D6}$ system. Setting to
zero the charges $q_2,q_3,q_4$ of the three $D2$ branes, one indeed
obtains the following $D6$---$\overline{D6}$ configuration,
\beq
ds_{10}^2= T_1^{1/8} \left(-dt^2+\sum_{i=1}^6 dx_i^2 \right)+
  T_1^{1/8}  \Sigma_1 \left({dr^2\over \Delta}+d\theta^2 \right)  
+\frac{\Delta \sin^2\theta}{T_1^{7/8}}d\varphi^2,
\labell{66bar}\eeq
which coincides with the metric constructed following \cite{dggh2,sen}
once the radial variable is shifted from $r \rightarrow r-M$ and the
string frame is used (see \eg \cite{youm} for the explicit expression).
Note that compactification on a $T^6$ torus yields
the metric, 
\beq
ds_{4}^2= T_1^{1/2} \left[-dt^2+\Sigma_1 \left({dr^2\over
\Delta}+d\theta^2 \right)\right] 
+\frac{\Delta \sin^2\theta}{T_1^{1/2}}d\varphi^2,
\eeq
describing the Einstein-Maxwell dilatonic single charge dihole
\reef{dilholem} when the coupling to the dilaton is $\alpha=\sqrt{3}$.
Also, \reef{66bar} may be uplifted to eleven dimensions to obtain
\beq
ds_{11}^2= -dt^2+\sum_{i=1}^6 dx_i^2 +
\Sigma_1\left({dr^2\over \Delta}+d\theta^2 \right) +
T_1\left(dx_{11}-2A_{\varphi}d\varphi\right)^2 
+\frac{\Delta \sin^2\theta}{T_1}d\varphi^2,
\labell{KK11}
\eeq
where the magnetic potential is given by,
\beq
A_{\varphi}=\frac{a_1M_1(r+M_1)\sin^2\theta}{\Delta+a_1^2\sin^2\theta}.
\eeq
After shifting $r \rightarrow r-M$ in the above metric, one exactly
recovers the Gross-Perry Kaluza-Klein dipole \cite{grossperry} embedded
in eleven dimensions, which is the starting point of Sen's analysis of
the $D6$---$\overline{D6}$ system. 

We would like to stress at this point that, instead of performing a
straightforward Kaluza-Klein compactification along the $x_{11}$
direction in \reef{KK11} to recover the configuration
\reef{66bar} (which possesses conical singularities along the axis), one
may reduce along a twisted direction \cite{dggh2,sen}.
In this way one obtains a configuration of $D6$---$\overline{D6}$ branes
suspended in a magnetic field. Precisely the same result is obtained if
a Harrison transformation with the appropriate value of the dilaton
coupling is performed directly on the reduced solution. As a matter of
fact, the equivalence between twisted KK reductions and Harrison
transformations in the reduced KK theory was proven in
\cite{dgkt93}. In the case at hand, we know that the effect of
performing a Harrison transformation on the metric \reef{66bar} is just
to replace $T_1$ by $\Lambda_1$ in \reef{66bar}. 

\subsection{The strings and membranes stretched between branes and
antibranes}

The proper length of a string stretched between the poles where branes
intersect depends in an essential way on the number of branes that
intersect. These strings stretch along the line $r=\gamma$, parametrized
by $\theta$, $0<\theta<\pi$. If there are less than four branes at the
intersection then the proper spatial distance between poles is finite,
but if all four branes are present then this distance is infinite. A
situation where things can be studied further is that where a IIA
configuration can be uplifted to $D=11$ supergravity. The line
$r=\gamma$ is fibered with the extra dimensions and becomes a surface.
As a consequence, the string stretching between branes becomes a
membrane. For the case of the $D6$---$\overline{D6}$ the study of such a
membrane was carried out in \cite{sen}. The configuration in (3) in the
previous subsection is also suitable for such an analysis, and will
allow us to recover as a particular case the results of \cite{sen}.

When we uplift \reef{uplift3} to eleven dimensions we obtain a
Kaluza-Klein dipole superposed to a system of three
intersecting delocalized $M2$ branes and three intersecting delocalized
$M2$ anti-branes, that is, 
\begin{eqnarray}
ds_{11}^2&=&  \left(T_2 T_3 T_4\right)^{2/3} \left[-dt^2 + 
\frac{ \Sigma_1\Sigma_2 \Sigma_3 \Sigma_4}{\left(r^2-\g\ct\right)^3} 
\left({dr^2\over \Delta}+d\theta^2 \right)\right]\nonumber\\
&+&\frac{\Delta \sin^2\theta}{T_1\left( T_2 T_3 T_4\right)^{1/3}}
d\varphi^2 + \frac{T_1}{\left(T_2T_3T_4\right)^{1/3}}\left(
dx_{11}-2A_{\varphi}d\varphi\right)^2\nonumber\\
&+& \frac{T_2^{2/3}}{T_3^{1/3}T_4^{1/3}}(dx_1^2 + dx_2^2)
+ \frac{T_3^{2/3}}{T_2^{1/3}T_4^{1/3}}(dx_3^2 +dx_4^2) 
+ \frac{T_4^{2/3}}{T_2^{1/3}T_3^{1/3}}(dx_5^2+dx_6^2).
\labell{uplift11}\end{eqnarray}
In M-theory, the open string state should be described by a membrane
wrapped on the surface $r=\gamma$. In order to distinguish it from the
other, self-gravitating M2-branes in the configuration, we will denote
this one as an $m2$-brane. As we just said, it wraps the surface
$r=\gamma$. This surface is a bolt of the Killing vector
${\bf q}=\partial/\partial x_{11}+
\frac{a_1}{2M_1(M_1+\gamma)}\partial/\partial\varphi$. It is convenient
to work with an ``adapted'' coordinate $\phi =
\varphi - \frac{a_1}{2M_1(M_1+\gamma)}x_{11}$ such that ${\bf q}\phi=0$.
In terms of this coordinate the metric induced on the bolt is given by
\begin{eqnarray}
ds_B^2&=& 
\prod_{i=2}^4 \left[
a_i^4\left(\frac{2q_i}{\gamma}+\frac{a_i^2\sin^2\theta}{\gamma
^2}\right)\right]
^{1/3}
\frac{(\gamma+M_1)^2-a_1^2\cos^2\theta}{\gamma ^4
\sin^2\theta}d\theta^2\nonumber\\
&+&\prod_{i=2}^4\left[
\frac{\g}{a_i^2}\left(\frac{2q_i}{\gamma}+\frac{a_i^2\sin^2\theta}
{\gamma ^2}\right)\right]^{1/3}
\frac{4M_1^2(\gamma+M_1)^2}{(\gamma+M_1)^2-a_1^2\cos^2\theta}d\phi^2,
\labell{bolt}
\end{eqnarray}
Now, when all four charges are turned on, this surface is topologically
a cylinder. Its shape, and therefore that of the $m2$-brane that wraps
it, is like a sphere
with two infinite funnels at its poles. This is most easily understood by 
looking at the geometry near the poles in $D=11$, after changing to the
coordinates \reef{change}. We know that near the poles the geometry is,
up to some angular distortion, the same as that of the core of an
intersection between a KK monopole and three M2-branes. But the latter
is
\begin{eqnarray}
ds_{11}^2&=& -dt^2+\sum_{i=1}^6 dx_i^2 +
{(q_2q_3q_4)^{1/3}\over q_1}\left(dx_{11}+q_1 \cos\bar\theta
d\varphi\right)^2\nonumber\\
&&+(q_2q_3q_4)^{1/3}q_1 \left({d\rho^2\over
\rho^2}+d\bar\theta^2+\sin^2\bar\theta d\varphi^2\right).
\end{eqnarray}
and we explicitly see that $\rho=0$ is down an infinite funnel of
constant curvature fibered with $x_{11}$
(there will be some angular distortion in the
situation at hand, though).

The proper area of the bolt
\begin{eqnarray}
A&=& \int d\theta d\phi\sqrt{g_{\theta\theta}g_{\phi\phi}}\nonumber\\
&=&2M_1(\gamma+M_1) \int  d\theta d\phi\prod_{i=2}^4\left(
\frac{2a_iq_i}{\gamma ^2}+\frac{a_i^3\sin^2\theta}{\gamma ^3}
\right)^{1/3}\frac{1}{\sin\theta},
\labell{area}
\end{eqnarray}
is infinite due to the divergence of the integration at $\theta=0,\pi$.
Since the energy of the $m2$-brane is $E=T_{m2} A$, where
$T_{m2}$ is the membrane tension, we reach the conclusion that the
energy of the $m2$-brane stretched inbetween the poles is infinite!
Notice that it remains infinite even if we set $a_1=0$ (so that $\gamma=M_1$).
If the latter is to be considered as the limit of coincidence of the
branes and antibranes, then the conclusion is even more striking than
that reached in \cite{sen}.

The situation, however, is different if one sets one, two or three $M2$
charges to zero in \reef{bolt}. Say that $n$ of these charges
are different from zero. Then, neglecting the angular distortion, the
geometry near the poles
is 
\begin{eqnarray}
ds_{11}^2&=& -dt^2+\sum_{i=1}^6 dx_i^2 +
{\tilde q^{1/3}\over q_1}\rho^{1-n/3}\left(dx_{11}
+q_1\cos\bar\theta
d\varphi\right)^2\nonumber\\
&&+\tilde q^{1/3}q_1 \rho^{-1-n/3}\left(d\rho^2
+\rho^2(d\bar\theta^2+\sin^2\bar\theta d\varphi^2)\right).
\end{eqnarray}
where $\tilde q$ is the product of the non-zero M2 charges.
It is straightforward to see that if $n\neq 0,3$, the geometry is
singular. When $n=0$ the geometry is just ${\bf R}^{10,1}$, since this is
the core of a KK monopole. The singularity for $n=1,2$ is nevertheless a
finite spatial distance away, so
the $m2$-brane is topologically a sphere, with curvature singularities
at the poles. Its proper area $A$, and the energy
of the $m2$-brane, is finite. 

Let us now consider the limit where the branes and antibranes are
coincident. It is not obvious what choice of parameters in the solution
should correspond to this limit. One would certainly require any
external fields to be absent, since their effect is to pull apart the
poles. For the $D6$---$\overline{D6}$ system it was argued in \cite{sen}
that one should also require $a=0$, since this minimizes the distance
between poles. In the present case, if we set all the $a_i$ to zero then
we would have all $M_i$ equal to each other, and as a consequence all
charges would be equal, which seems too restrictive. In order to
motivate other alternatives, notice first that all the branes at an
intersection move together, since they all intersect at the same pole.
Then, it might be enough to set just one of the $a_i$ to zero. This
would leave us with four parameters, which can be regarded as the four
charges of the branes, as desired. Since $a_1$ is singled out as
characterizing the twist in the eleventh direction, then, at least in
the context of M-theory, it probably makes more sense to define the
coincidence limit by setting only $a_1=0$ (so that $\gamma=M_1$), while
leaving $a_2,a_3,a_4$ arbitrary. 

With this choice, the conclusions we have reached above for the m2-brane
stretched between branes and antibranes still hold in the coincidence
limit: when any of the charges of the M2-branes is zero the proper area
of the m2-brane is finite, whereas it is infinite when all charges are turned
on. This conclusion, which may be taken as a `prediction' about the
strong coupling limit of this brane-antibrane configuration, is even
more striking than that reached in \cite{sen} for $D6$---$\overline{D6}$
branes.

\section{Discussion and Outlook}
\renewcommand{\theequation}{5.\arabic{equation}}
\setcounter{equation}{0}

Although BPS states in string/M-theory have been under very close
scrutiny over the last few years, they are far from providing us with
the full spectrum of such theories. For instance, Sen has recently
highlighted the r\^ole played by D-brane---anti-D-brane systems in
constructing both unstable and stable non-BPS D-branes \cite{sen99}.
Motivated by his earlier work \cite{sen} on a $D6$---$\overline{D6}$
configuration of type IIA string theory, which he relates to the
Kaluza-Klein dipole solution \cite{grossperry} (Euclidean 4d Kerr
metric) embedded in eleven dimensions, and also by the recent work of
one of us \cite{blackdih} on black diholes, we have identified new
classical exact solutions to 4d General Relativity containing $n$
abelian gauge fields and $n-1$ independent scalar fields, whose generic
lagrangians are given by \reef{unaction}. As already noted in the
introduction, the case of two electric and two magnetic charges ($n=4$)
is particularly interesting since the corresponding lagrangian
\reef{straction} arises as a consistent truncation of a wide class of
low energy superstring compactifications, and therefore, the new 4d
solutions can be uplifted and interpreted as rather sophisticated
brane---anti-brane systems.

For general values of $n$, our 4d solutions depend on $n+1$ parameters
$\{\gamma, a_i, i=1,\ldots n\}$, are static, axisymmetric and describe
composite diholes. Indeed, the near horizon analysis of these solutions
reveals they contain two throats which one can identify, for arbitrary
values of the parameters (which label the $n$ charges and the separation
between the holes), with the throats of two oppositely charged extremal
composite black holes. Although the composite dihole configurations are
plagued with conical singularities, it is possible to suspend them in
external magnetic fields via generalised Harrison transformations, and
tune these fields to values which eliminate the conical deficit and keep
the configurations in equilibrium, though unstable: a slight deviation
along the symmetry axis from the equilibrium configuration is expected
\cite{dggh2} to result in either making the composite black holes
collapse onto each other or accelerate apart. This instability, however,
is of a completely different nature from the tachyonic instability
appearing in perturbative string theory \cite{sen99}. One might even
wonder if the latter is a feature that survives when the string coupling
is increased and the effects of self-gravity become important.

The total charge of the $U(1)^4$ composite diholes is zero while their
ADM mass, whether or not they are suspended in external fields, is given
by $E=\frac{1}{2} \sum_{i=1}^4 M_i=\frac{1}{2} \sum_i
(\gamma^2-a_i^2)^{1/2}$ and is generically 
strictly positive.\footnote{We are not considering the possibility of
diholes made out of `massless' black holes.} These configurations are
therefore non extremal, and when analysed in a context of supersymmetry,
break all supersymmetries, an observation which is particularly obvious
in the presence of external fields, which are asymptotically Melvin and
therefore have no Killing spinors associated to them. Another
interesting observation is that the mass of the two composite extremal
black holes, which is equal to $2M_{bh}=\frac{1}{2}\sum_{i=1}^4
\frac{M_i(M_i+\gamma)}{\gamma}$, exceeds the ADM mass $E$ of the
composite dihole: the latter is therefore non-marginally bound. When
uplifted to ten or eleven dimensions, it becomes a supergravity soliton
and can be interpreted as a system of four intersecting branes and four
intersecting anti-branes as sketched in Figure 1, with the (anti)branes
localized in the overall transverse directions only. Each brane is
charged under a different $U(1)$ and has its corresponding anti-brane
parallel to it. The existence of such configurations of branes and
antibranes at arbitrary (large) separation $2\gamma$ when $a_i \gg M_i$ is
another indication that the static force vanishes between the branes and
antibranes.

We thus succeeded in providing classical solutions to supergravity
theories which describe static, zero charge configurations which appear
as a cluster of intersecting charged branes and a cluster of
intersecting charged anti-branes. They are non-trivial generalisations
of the $D6$---$\overline{D6}$ systems analysed by Sen, who recognised
that systems with {\it coincident} branes and anti-branes could in
particular be used in the construction of stable non-BPS states of a new
type, via orientifolding and orbifolding. His analysis relies on the
perturbative description of D-branes, and it would be very instructive
to study whether one can deform the supergravity solitons associated to
brane---anti-brane systems in such a way that the resulting solutions
(if any) describe these new stable non-BPS states. A first step in this
direction should involve the study of the coincidence limit of the
branes and antibranes in configurations of the sort we have been
discussing. However, it appears that in the presence of gravity these
systems exhibit features markedly different to those seen at the
perturbative level, in particular, the size of membranes (and strings)
stretched between the branes and antibranes remains non-zero (even
infinite) in the limit of coincidence. This may cast some doubt on the
possibility of carrying out comparisons between calculations performed
at the weakly (CFT) and strongly coupled (supergravity) regimes.

\section*{Acknowledgements}

We would like to thank Ruth Gregory and Simon Ross for conversations and
comments on the manuscript. AC is supported by the Royal Thai Government
through DPST scholarship. RE is supported by EPSRC through grant
GR/L38158 (UK), and by grant UPV 063.310-EB187/98 (Spain). AT
acknowledges the Leverhulme Trust for a Fellowship.

\end{document}